\documentclass[aps,prb,reprint,groupedaddress,footinbib,superscriptaddress]{revtex4-1}
\usepackage{amsmath}
\usepackage{graphicx}
\usepackage{hyperref}
\usepackage{bbold}
\usepackage{xcolor}
\usepackage[normalem]{ulem}

\begin{document}
	\title{Current rectification in junctions with spin-split superconductors}
\author{Stefan Ili\'{c}}
\affiliation{Centro de F\'{i}sica de Materiales (CFM-MPC), Centro Mixto CSIC-UPV/EHU, 20018 Donostia-San Sebasti\'{a}n, Spain} 

\author{P. Virtanen}
\affiliation{Department of Physics and Nanoscience Center, University of Jyväskylä, P.O. Box 35 (YFL), FI-40014 University of Jyväskylä, Finland}

\author{T. T. Heikkilä}
\affiliation{Department of Physics and Nanoscience Center, University of Jyväskylä, P.O. Box 35 (YFL), FI-40014 University of Jyväskylä, Finland}

\author{F. Sebasti\'an Bergeret}
\affiliation{Centro de F\'{i}sica de Materiales (CFM-MPC), Centro Mixto CSIC-UPV/EHU, 20018 Donostia-San Sebasti\'{a}n, Spain}
\affiliation{Donostia International Physics Center (DIPC), 20018 Donostia-San Sebasti\'{a}n, Spain}

\begin{abstract}
Spin-split superconductors exhibit an electron-hole asymmetric spin-resolved density of states, but the symmetry is restored upon averaging over spin. On the other hand, asymmetry appears again in tunneling junctions of spin-split superconductors with a spin-polarized barrier. As demonstrated recently in both theory and experiment, this fact leads to a particularly strong thermoelectric effect in superconductor-ferromagnet structures. In this work we show another important effect stemming from the electron-hole asymmetry - current rectification. We calculate the charge current in spin-polarized tunnel junctions of normal metal and a spin-split superconductor with AC and DC voltage bias. In the DC case, the I-V curve is not fully antisymmetric and has a voltage-symmetric component due to spin polarization. This translates to the existence of a rectified current in the AC case, which is proportional to the spin polarization of the junction and strongly depends on the frequency of the applied bias.  We discuss possible applications of the rectification effect, including a diode for superconducting electronics and radiation detectors. The analysis of the rectified charge current is supplemented by the discussion of heat current and relevant noise correlators, where electron-hole asymmetry also plays an important role, and  which are useful for applications in detectors. 
\end{abstract}
\maketitle

\section{Introduction}
\begin{figure}[h!]
	\includegraphics[width=0.48\textwidth]{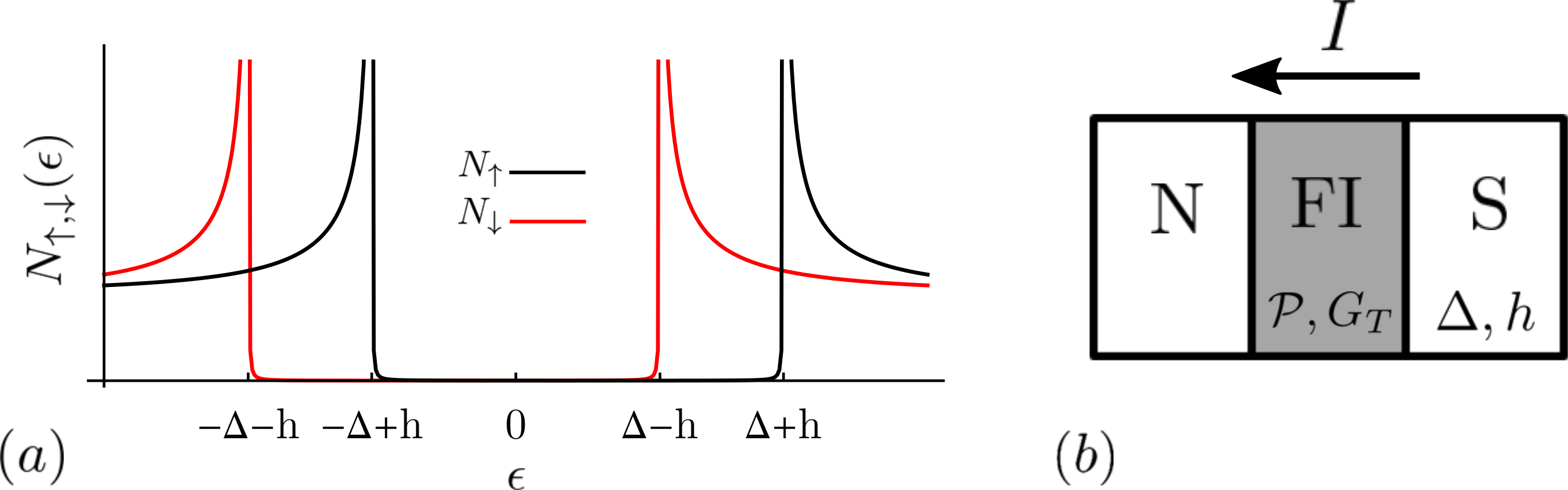}
	\caption{\label{fig1}$(a)$ Density of states for spin-up ($N_\uparrow$) and spin-down ($N_\downarrow$) in a spin-split superconductor. $\Delta$ is the superconducting gap, and $h$ is the exchange field. $(b)$ Schematic representation of the N-FI-S junction. FI layer acts as a tunneling barrier with a conductance $G_T$ and spin-polarization $\mathcal{P}$. The arrow denotes the direction of the rectified current when $h \mathcal{P}>0$ (the direction is opposite for $h \mathcal{P}<0$).}
\end{figure}

Superconductors in the presence of an exchange field become spin-split, with a different density of states (DoS) for spin-up and spin-down quasiparticles (see Fig.~\ref{fig1}). Spin-split superconductors have been intensively studied for several decades already \cite{meservey_tunneling_1975}, but lately the interest in them has renewed, as they have been identitfied as an important ingredient for many applications including superconducting spintronics \cite{ bergeret_odd_2005, linder_superconducting_2015, bergeret2018colloquium}, topological superconductivity \cite{leijnse_introduction_2012, sato_topological_2017}, and radiation detectors \cite{bergeret2018colloquium, heikkila2019thermal}.

Spin-split DoS can be achieved by applying an external in-plane magnetic field to a thin superconductor\cite{meservey_tunneling_1975}, but this approach requires strong magnets and precise alignment of the field in order to avoid suppression of superconductivity by the orbital effect \cite{tinkham2004introduction}. Alternative way, which is more suitable for applications, is to attach a magnetic material to the superconducor. In  this way the magnetic proximity effect  leads to  an exchange field in the superconductor \cite{bergeret_inverse_2005}.
The most promising structures of this type are junctions of ferromagnetic insulator (FI) and a superconductor (S) \cite{tokuyasu_proximity_1988, heikkila2019thermal}. Well defined and sharp spin-splitting was demonstrated in numerous experiments in which S is an aluminum layer, and FI is a europium chalcogenide (EuO, EuS, EuSe) \cite{tedrow_spin-polarized_1986, moodera_electron-spin_1988, hao_spin-filter_1990, hao_thin-film_1991, meservey_spin-polarized_1994, moodera_phenomena_2007, xiong_spin-resolved_2011, strambini_revealing_2017, rouco2019charge}, even in the absence of externally applied magnetic fields. The induced exchange field in S is homogeneous if it is thinner than the coherence length \cite{hijano_coexistence_2021}. 

As illustrated in Fig.~\ref{fig1}$(a)$, the spin-resolved DoS $N_s(E)$ ($s=\uparrow, \downarrow$, $E$ is the energy) in a spin-split superconductor  is strongly electron-hole (e-h) asymmetric, $N_s(E)\neq N_s(-E)$. The symmetry is, however,  restored in total density of states:  $N_\uparrow(E)+N_\downarrow(E)=N_\uparrow(-E)+N_\downarrow(-E)$. On the other hand, the asymmetry appears again if the spin-split superconductor is incorporated in a junction with some spin-filtering element. For instance, in FI-S junctions, in addition to being a source of the spin-splitting field, FI can act also as a spin-filtering tunneling barrier \cite{heikkila2019thermal}. This is demonstrated in several experiments with FI-S junctions \cite{tedrow_spin-polarized_1986, hao_spin-filter_1990, senapati_spin-filter_2011, moodera_electron-spin_1988, rouco2019charge}, where the measured differential conductance is not fully symmetric in voltage due to the combination of spin-splitting and spin-filtering. 

One important consequence of e-h symmetry breaking is coupling of charge transport and temperature gradients, leading to thermoelectric effects, as is well established in conventional semiconducting structures. Similar effects were predicted in N-FI-S junctions \cite{ozaeta_predicted_2014,machon_nonlocal_2013} (N is a normal metal), and recently demonstrated in experiment \cite{kolenda_observation_2016, kolenda_thermoelectric_2017}. This effect can be exploited to engineer an ultra-sensitive THz radiation detector based on superconducting junctions \cite{heikkila_thermoelectric_2018}.

The main focus of this work is another manifestation of e-h symmetry breaking in N-FI-S junctions - non-reciprocal charge transport, or the diode effect. This means that in the case of DC bias, the I-V characteristc of the junction is not fully antisymmetric - $I(V)\neq -I(-V)$ and thus one current direction is preferred. Consequentially, AC currents will be rectified. Therefore, N-FI-S junctions act as a current rectifier - an element with numerous possible applications in cryogenic and superconducting electronics and spintronics. In contrast to conventional semiconducting diodes, a N-FI-S diode operates at much lower temperatures and voltages, and therefore has much less dissipation.  Moreover, the rectification effect in these junctions can be used together with thermoelectric effect (or instead of it) in the realization of the above-mentioned superconducting radiation detectors.  

The realization of a superconducting diode has been a particularly active topic in the past couple of years. Most theoretical \cite{wakatsuki2018nonreciprocal, he2021phenomenological} and experimental \cite{wakatsuki2017nonreciprocal, ando_observation_2020} efforts have focused on exploiting the so-called magnetochiral anomaly in systems with spin-orbit coupling and an applied magnetic field, which leads to a non-reciprocal critical current. This leads to rectification of \emph{supercurrents}, but only in a limited current range. N-FI-S junction is conceptually different, as it rectifies \emph{quasiparticle currents}  and operates at arbitrary currents.  Moreover, it does not require an external field.  

In addition to rectification of charge current, e-h symmetry breaking in N-FI-S has important consequences for other transport quantities - electronic heat currents and relevant noise correlators.  For instance, voltage-driven heat current has as a component antisymmetric in DC voltage bias, and odd-order harmonics in AC bias, both of which cannot exist if e-h symmetry is satisfied.  Moreover, in the presence of an electromagnetic environment, N-FI-S junctions will rectify the environmental noise, leading to a finite current even without a bias voltage or temperature gradient. These findings are particularly important for applications in detectors, where both charge and heat transport need to be taken into account, and  noise is a limiting factor for the performance. 

In this work we provide a complete theoretical characterization of charge and heat transport in N-FI-S junctions, including the noise, for both DC and AC voltage bias. We discuss the charge transport in Sec.~\ref{RectCharge}, heat transport in Sec.~\ref{Heat}, and analyze the noise in Sec.~\ref{Noise}. In Sec.~\ref{RectvsTherm}, we discuss the interplay of the rectification and thermoelectric effects in N-FI-S junctions, which has important implications for possible applications of these structures in radiation detection. Finally, in Sec.~\ref{DCB} we investigate the junction in the presence of an electromagnetic environment. For brevity, in the main text we discuss only the main results of our calculations. Additional details are presented in Appendices \ref{app1} and \ref{app2}, where we derive all our results using  the non-equilibrium Keldysh Green's function formalism \cite{kopnin2001theory, bergeret_odd_2005}.

In this work we primarily focus on the N-FI-S junction, shown in Fig.~\ref{fig1}$(b)$, in which FI acts both as a source of the exchange field in the superconductor, and the spin-filtering barrier. However, note that all our results should also hold in any superconducting structure where both spin-splitting and spin-filtering are present, so that the e-h symmetry is broken. 

\section{Rectification of the charge current \label{RectCharge}}
In this section, we analyze the charge current in N-FI-S junctions. We start from the DC biased case in Sec.~\ref{SecIDC}, where we show that the I-V characteristic is not fully antisymmetric due to e-h symmetry breaking.  We proceed to analyze the AC-biased case in Sec.~\ref{SecIAC}, where we show the existence of a rectified current, which is proportional to the spin-polarization of the junction. Derivations of main expressions from this section can be found in Appendix \ref{app1}. 
\subsection{DC case \label{SecIDC}}
 The  current flowing through the N-FI-S junction in the DC bias is given by the tunneling formula \cite{ozaeta_predicted_2014, giazotto_ferromagnetic-insulator-based_2015}
\begin{equation}
	I=\frac{G_T}{e}\int dE [N_++\mathcal{P}N_-][f_N(\tilde{E})-f_S(E)].
	\label{eqIDC}
	\end{equation}
Here $G_T$ is the tunneling conductance, and $f_{N,S}(E)=[1+\exp(E/T_{N,S})]^{-1}$ are the Fermi functions, with $T_N$ and $T_S$ being the temperatures of the normal and superconducting side of the junction, respectively. We introduced $\tilde{E}=E-eV_{dc}$, where $V_{dc}$ is the voltage bias, and   $N_{\pm}(E)=\frac{1}{2}[N_{\uparrow}(E)\pm N_\downarrow(E)]$. $N_+$ is the total DoS of the superconductor, whereas $N_-$ corresponds to the difference of DoS for spin-up and -down, and therefore it only exists in a spin-split superconductor. Finally, the parameter $\mathcal{P}\in [0,1]$ describes the spin-polarization of the junction, where $\mathcal{P}=0$ corresponds to absence of polarization, and $\mathcal{P}=1$ is perfect polarization. 
In the following, we assume that the DoS is given by the spin-split BCS expression $N_{\uparrow,\downarrow}(E)= \text{Re}\bigg[(E\pm h+i\eta)/\sqrt{(E\pm h+i\eta)^2-\Delta^2}\bigg]$, where $\Delta$ is the superconducting gap, $h$ is the exchange field responsible for spin-splitting, and $\eta$ is the so-called Dynes parameter \cite{dynes_tunneling_1984}, which accounts for inelastic scattering. Note that we ignore spin-flip and spin-orbit scattering, which are expected to be weak in many experimentally available FI-S structures \cite{rouco2019charge, hijano_coexistence_2021}. Strong disorder of this kind would introduce significant ``smearing" to the DoS, thus reducing the asymmetry of the I-V curve, in turn negatively impacting the rectification properties of the N-FI-S junction.

\begin{figure}[h!]
	\includegraphics[width=0.48\textwidth]{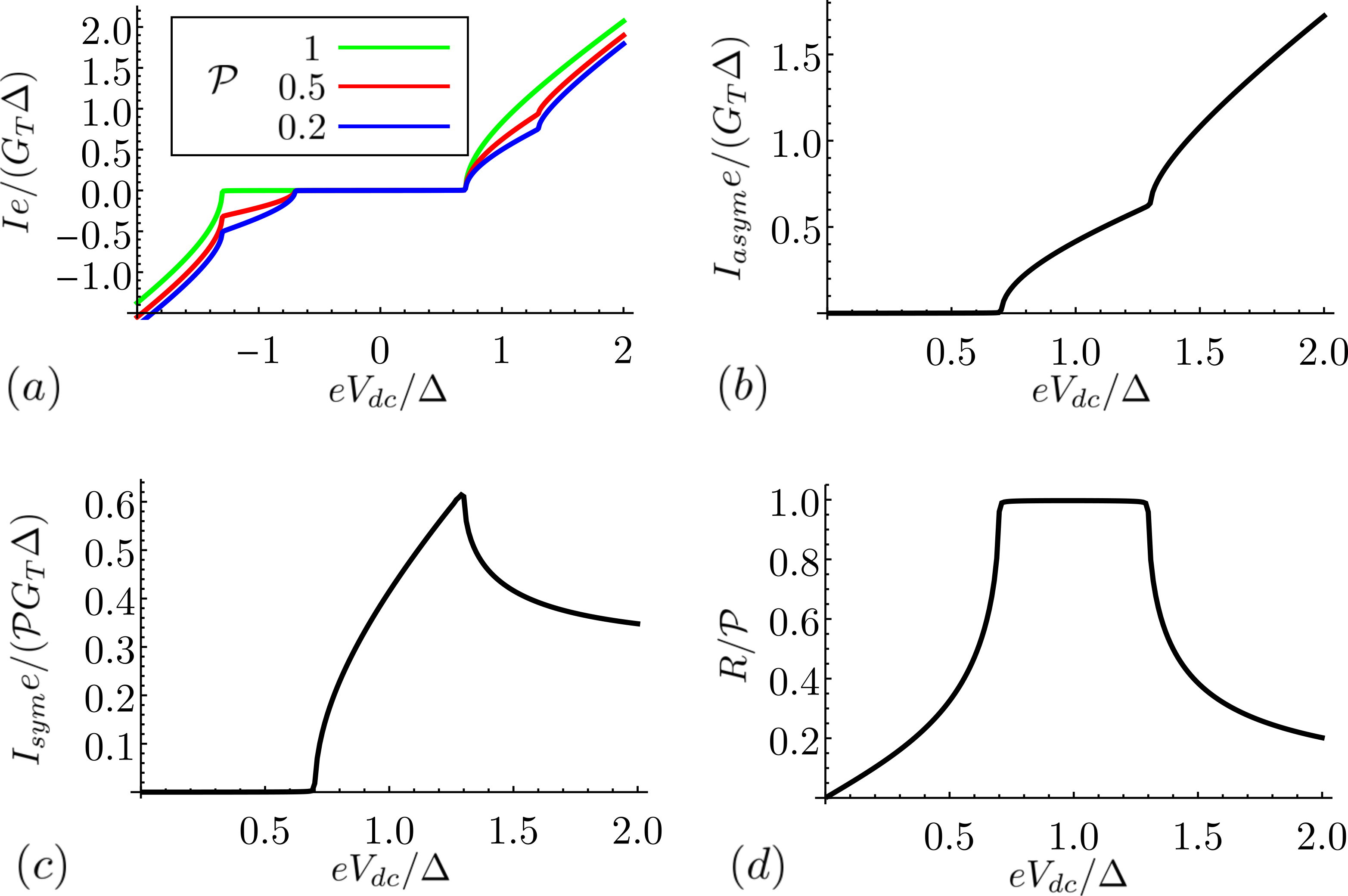}
	\caption{\label{figDC} Current in  N-FI-S junctions with applied DC bias $V_{dc}$ at $T_N=T_S=0$. $(a)$ I-V curve for different values of the polarization $\mathcal{P}$. $(b)$ Voltage-asymmetric component of the current. $(c)$ Voltage-symmetric component of the current. $(d)$ Rectification coefficient. The following parameters were used in all plots: $h=0.3 \Delta$, $\eta=10^{-3} \Delta$.}
\end{figure}

First, let us analyze Eq.~\eqref{eqIDC} in the limit of very low temperature ($\frac{\eta}{\Delta}\gg e^{-\Delta/T}$, with $\eta, T\ll \Delta$). We plot the current in Fig.~\ref{figDC}$(a)$. To illustrate the non-reciprocity of the current, we split it into the symmetric component $I_{sym}=\frac{1}{2}|I(V_{dc})+I(-V_{dc})|$, and the antisymmetric component $I_{asym}=\frac{1}{2}|I(V_{dc})-I(-V_{dc})|$, which are plotted in Figs.~\ref{figDC}$(b)$ and $(c)$, respectively. $I_{asym}$ stems from the $N_+$ contribution in Eq.~\eqref{eqIDC} - this is the  conventional component. The more important component for our discussion is  $I_{sym}$, which stems from  $\mathcal{P}N_-$ in Eq.~\eqref{eqIDC}, and therefore it can only exist if both spin-filtering and spin-polarization are present. This component determines the rectification capabilities of the junction, i.e. conversion of an  AC to a  DC current. This conversion  can be quantified by the rectification coefficient $R$
\begin{equation}
R=\frac{I_{sym}}{I_{asym}}.
\end{equation}
As shown in Fig.~\ref{figDC}, the rectification coefficient plateaus in the voltage window $eV_{dc}\in [\Delta-h,\Delta+h]$, and its maximum value is determined by the junction polarization $\mathcal{P}$.

Let us examine more closely the low-voltage regime, $eV_{dc}<\Delta-h$. In the low temperature limit $\frac{\eta}{\Delta}\ll e^{-\Delta/T}$, all current stems from the subgap states introduced by inelastic scattering. For a weak Dynes parameter  $\eta \ll \Delta-h$, we find the following expression for the current
\begin{equation}
I=\eta G_T [F_{asym}(eV_{dc},h)+\mathcal{P} F_{sym}(eV_{dc},h)].
\label{eqDynes}
\end{equation}
Here we introduced the functions $F_{asym}(eV_{dc},h)=\frac{1}{2}[F(eV_{dc}+h)+F(eV_{dc}-h)]$, $F_{sym}(eV_{dc},h)=\frac{1}{2}[F(eV_{dc}+h)-F(eV_{dc}-h)-2F(h)]$, with $F(x)=x/\sqrt{\Delta^2-x^2}$. The rectification coefficient is then
\begin{equation}
R=\mathcal{P}\frac{F_{sym}}{F_{asym}}.
\label{eqRDynes}
\end{equation}
Equation \eqref{eqRDynes} describes the low-voltage part ($eV_{dc}<\Delta-h$) of the curve shown in Fig.~\ref{figDC}$(d)$.  
In the limit $h,eV_{dc}\ll \Delta$, we may approximate $I \approx \frac{\eta eV_{dc}}{\Delta}[1+\frac{3}{2}\frac{eV_{dc} \mathcal{P} h}{\Delta^2}]$, and therefore $R\approx \frac{3 \mathcal{P}}{2}\frac{eV_{dc}h}{\Delta^2}$.

\begin{figure}[h!]
\includegraphics[width=0.5\textwidth]{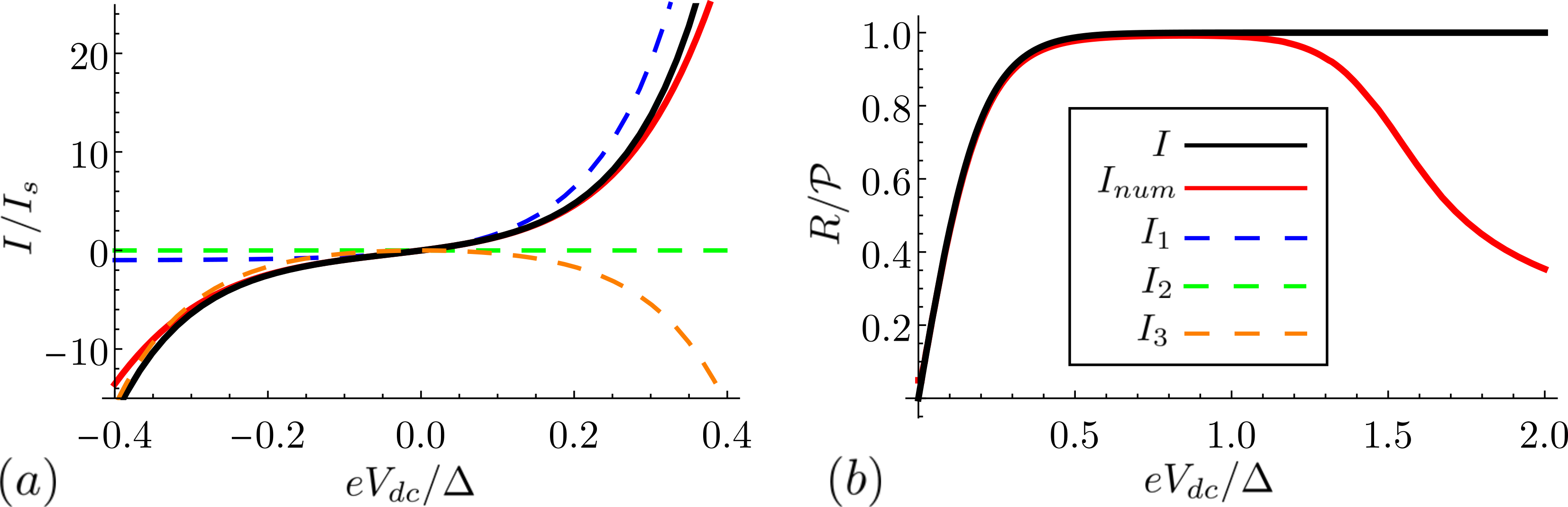}
\caption{\label{figDiode} Comparison between  the results obtained  from the diode equation and the exact numerics. Panel $(a)$ shows  the current and panel $(b)$ the rectification coefficient. Full black curves correspond to analytical expressions given by Eqs.~\eqref{eqDiode} and \eqref{eq:R}. Red curve is obtained by numerically evaluating energy integrals in Eq.~\eqref{eqIDC}. Dashed curves correspond to different current components $I_{i}$ that enter  Eq.~\eqref{eqDiode}. Parameters used in the plot: $h=0.5 \Delta$, $\mathcal{P}=0.4$, $T=0.1 \Delta$, $\eta=10^{-4}\Delta$. }
\end{figure}

 Now, let us turn to a regime of higher temperatures $e^{-\frac{\Delta}{T}}\gg \frac{\eta}{\Delta}$, with $\Delta\gg \eta$. At low voltages $eV_{dc}<\Delta-h$, we find that the current can be written in a form of a non-ideal Shockley diode equation \cite{shockley1949theory}. Namely, the current has three components $I=I_1+I_2+I_3$ given as:
\begin{align}
&I_1=I_S\left(e^{eV_{dc}/T}-1\right), \quad I_2=I_S e^{-2h/T}  \left(1-e^{-eV_{dc}/T}\right), \nonumber \\
&I_3=I_S\left(1-e^{-2h/T}\right) \left[\cosh\left(\frac{eV_{dc}}{T}\right)-1\right](\mathcal{P}-1),
\label{eqDiode}
\end{align}
where $ I_S \equiv \frac{G_T}{e} \Delta K_1\left(\frac{\Delta}{T}\right)e^{h/T}$, and $K_1$ is the Bessel K function. The component $I_1$ corresponds to the ideal Shockley contribution, whereas $I_2$ and $I_3$ describe deviations from the ideal behavior. Importantly,  $I_2$ vanishes if $h\gg T$, whereas $I_3$ vanishes if polarization is ideal, $\mathcal{P}=1$. Therefore, if both these conditions are met, N-FI-S junctions behave as ideal Shockley diodes.  In Fig.~\ref{figDiode}$(a)$, we plot the different components given by Eq.~\eqref{eqDiode}, and compare the total current $I$ obtained this way with the exact numerical solution.

Using Eq.~\eqref{eqDiode}, we find that the rectification coefficient is
\begin{equation}
    R = \mathcal{P} \tanh\left(\frac{h}{T}\right) \tanh\left(\frac{eV_{dc}}{2T}\right).
    \label{eq:R}
\end{equation}
If the ideality conditions are met, $h\gg T$ and $\mathcal{P}=1$, Eq.~\eqref{eq:R} reduces to $R=\tanh\left(\frac{eV_{dc}}{2T}\right).$ In Fig.~\ref{figDiode}$(b)$, we plot Eq.~\eqref{eq:R} and compare it to the exact numerical solution. 

Let us summarize the discussion above. Depending on the temperature, we found two regimes of rectification. At very low temperatures, $\frac{\eta}{\Delta}\gg e^{-\frac{\Delta}{T}}$, the low-voltage transport is dominated by the inelastic scattering, and Eq.~\eqref{eqRDynes} holds. On the other had, if the temperature is large-enough, $\frac{\eta}{\Delta}\ll e^{-\frac{\Delta}{T}}$, we find behavior similar to the Shockley diode equation, and Eq.~\eqref{eq:R} holds. Importantly, both regimes exhibit $R_{max}=\mathcal{P}$, and therefore the junction can serve as an equally good rectifier in the two regimes. However, note that $R_{max}$ is reached at different voltages: at $eV_{dc}\sim \Delta-h$ in the first regime, and at $eV_{dc}\sim T$ in the second one.

\begin{figure}[h!]
\includegraphics[width=0.3\textwidth]{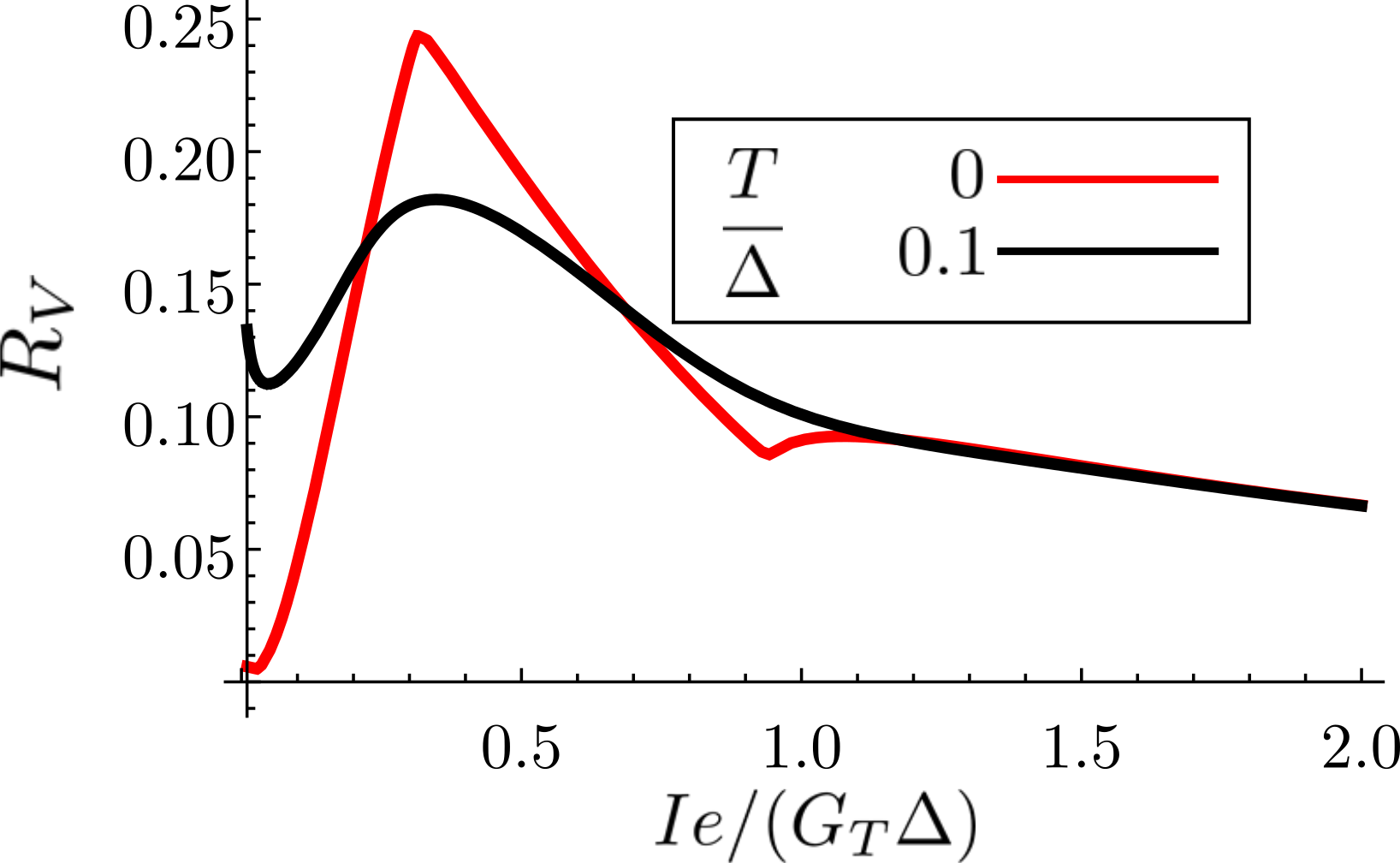}
\caption{\label{FigIbias}Rectification ratio $V_{sym}/V_{asym}$ in the current-biased regime. Parameters used: $h=0.3\Delta$, $\mathcal{P}=0.5$, $\eta=10^{-3}\Delta$.}
\end{figure}

 So far, we have discussed the voltage-bias regime. However, some  experimental situations corresponds to a  current bias regime. To check the rectification properties in this case we define the symmetrized and antisymmetrized voltage,  $V_{sym}=\frac{1}{2}|V(I)+V(-V)|$ and $V_{asym}=\frac{1}{2}|V(I)-V(-V)|$, and plot their ratio, $R_V$,  in Fig.~\ref{FigIbias} for different temperatures. We see that the rectification capabilities of the junction are still substantial in this regime, but significantly reduced in comparison with the voltage-biased case. Specifically, for the parameters used in Fig.~\ref{FigIbias}, we find $R_V=V_{sym}/V_{asym}<\mathcal{P}/2$.
 We may model the finite-temperature result ($\eta=0$) by inverting Eq.~\eqref{eqDiode} so that $V_{\rm dc}$ is a function of the current $I$. This yields the rectification coefficient 
 \begin{equation}
     R_V=-1+\frac{2}{1-\frac{\ln\left(\frac{\tilde I + P +\sqrt{1+\tilde I^2+2\tilde I P}}{1+P}\right)}{\ln\left(\frac{-\tilde I+P+\sqrt{1+\tilde I^2-2 \tilde I P}}{1+P}\right)}},
 \end{equation}
 where $\tilde I=I/I_S$. This has a maximum at $I \approx I_S$ and for $P\lesssim 0.9$ the maximum rectification is $R_V(I=I_S) \approx P/3$.

\subsection{AC case \label{SecIAC}}
In the AC voltage-biased case, the time-dependent current $I(t)$ can be written as a sum of harmonics
    \begin{equation}
I(t)=I_0+\sum_{m=1}^{\infty}\bigg[ I_m^c \cos (m  \omega t)+I_m^s \sin (m \omega t)\bigg].
\label{Eq:I(t)}
    \end{equation}
Here, $I_0$ is the time-averaged, or rectified, component, while $I_{m}^{s,c}$ are higher-order harmonics. The expression for the rectified current resembles the well-known Tien-Gordon formula for photo-assisted transport \cite{tien1963multiphoton, tucker1985quantum, virtanen2016stimulated}, namely
    \begin{equation}
I_0=\frac{G_T}{e}\sum_n J_n^2(\beta) \int \text{d}E\, \mathcal{P}N_- [f_N(E_n)-f_S(E)].
\label{Eq:I0}
\end{equation}
 Here, $J_n$ are the Bessel functions, $\beta=eV_{ac}/\omega$, where $V_{ac}$ is the amplitude of the AC voltage, and $\omega$ is its frequency, and we introduced $E_n=E+n\omega$. Higher-order harmonics are
\begin{multline}
I_m^c=\frac{G_T}{e}\sum_n \mathcal{M}_{mn}^+(\beta) \\ \int dE [f_N(E_n)-f_S(E)] 
\begin{cases}
\mathcal{P}N_-,  &\text{even }m  \\
N_+,  &\text{odd }m.
\end{cases}
\label{Eq:Ic}
\end{multline}
\begin{equation}
I_m^s=\frac{-iG_T}{e}\sum_n \mathcal{M}_{mn}^-(\beta) \int dE f_N(E_n) 
\begin{cases}
\mathcal{P}M_-,  &\text{even }m  \\
M_+,  &\text{odd }m.
\end{cases}
\label{Eq:Is}
\end{equation}
Here $\mathcal{M}_{nm}(\beta)=J_n(\beta)[J_{n+m}(\beta)\pm J_{n-m}(\beta)]$. Note that in Eq.~\eqref{Eq:Is} we introduced $M_{\pm}=\frac{1}{2}[M_\uparrow (E)\pm M_\downarrow (E)]$, with $M_{\uparrow,\downarrow}(E)= i\text{Im}\bigg[(E\pm h+i\eta)/\sqrt{(E\pm h+i\eta)^2-\Delta^2}\bigg]$. Here, the quantity $M$ comes from the so-called kinetic inductance of the junction. Unlike the density of states $N$, the quantity $M$ is finite inside the superconducting gap, and vanishes outside of it. 
As we see from Eqs.~\eqref{Eq:I0}-\eqref{Eq:Is}, the combination of finite spin-splitting and spin-filtering in the junction generates a rectified current $I_0$, as well as even harmonics $I_{2k}^c$ and $I_{2k}^s$.

\begin{figure*}[t!]
	\includegraphics[width=0.9\textwidth]{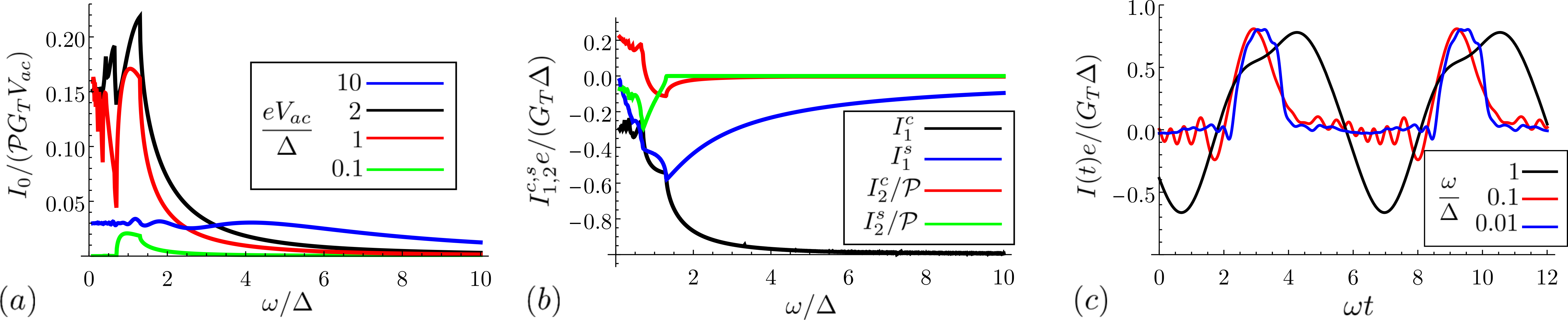}
	\caption{\label{figAC}Charge current in AC-biased N-FI-S junctions at $T_N=T_S=0$. In all plots we use the following parameters: $h=0.3 \Delta$, $\eta=10^{-3}\Delta$.  $(a)$ Time-averaged current as a function of frequency, at different amplitudes of the bias voltage. $(b)$ First and second harmonic of the charge current as a function of frequency, at $eV_{ac}=\Delta$. $(c)$ Time-dependent current, for various values of the frequency, at $\mathcal{P}=0.9$ and $eV_{ac}=\Delta$. Note that we choose a different normalization for the charge current in panel $(a)$,  and panels $(b)$ and $(c)$ in order to have a better visibility of the different curves. }
\end{figure*}

We plot $I_0$ as a function of frequency in Fig.~\ref{figAC}$(a)$. At a lower frequency, $I_0$ exhibits resonance-like features which come from photon-assisted tunneling processes, and decays at higher frequencies. In the high-frequency limit $\omega\gg \Delta, h, eV_{ac}\gg T$, we find
\begin{equation}
I_0=\frac{\mathcal{P} G_T}{2e}\frac{e^2V_{ac}^2}{\omega^2}h.
\end{equation}
This expression is used in Sec.~\ref{RectvsTherm}, where we discuss the implications of $I_0$ for radiation detectors based on N-FI-S junctions. Importantly, the direction of the rectified current in determined by the product $\mathcal{P} h$.\footnote{The sign of $h \mathcal{P}$  varies  from system to system. It depends on the material combination, quality of the interface, etc. $h$ stems from the interfacial exchange between the localized moments of the FI and the conduction electrons\cite{zhang2019theory} which participate in the superconducting state.  
In other words,  $h$ is uniquely determined by interfacial effects which takes place over atomic distances.  On the  other hand, the polarization $\mathcal{P}$ of the  tunnelling barrier is determined by the band structure  of the  FI layer.  In certain ideal cases, $h$ and $\mathcal{P}$ can be related \cite{tokuyasu_proximity_1988}, but for other  systems, as for example LCO-Pt\cite{velez2019spin}, the interfacial spins decouple magnetically from the rest of the  FI film and hence $h$ and $\mathcal{P}$ are independent parameters. 
}
At $\mathcal{P}h>0$,  $(\mathcal{P}h<0)$, this direction is from S to N (N to S), as shown in Fig.~\ref{fig1}$(b)$. Therefore, the direction of the N-FI-S diode can be changed by tuning the exchange field $h$, for instance by applying an external field, which could be particularly useful for various applications in cryogenic electronics and spintronics.

The first and second harmonics are plotted in Fig.~\ref{figAC}$(b)$. They are of comparable magnitude at low frequency, whereas at high frequencies $I_1^c$ dominates so that $I(t)\approx - G_T V_{ac}\cos(\omega t)$ (the standard Ohm's law for AC currents).
Finally, in Fig.~\ref{figAC}$(c)$, we plot the time dependent current using Eq.~\eqref{Eq:I(t)}.
Importantly, we see that at low frequencies $\omega\ll \Delta$ the junction behaves as an almost ideal half-wave rectifier. 

Rectification of AC signals has already been noticed in a recent experiment \cite{quay2016frequency}, in a structure consisting of a thin Al superconductor, which is spin-split by a Zeeman field, and a ferromagnet. The results of this section suggest a few interesting directions that could be explored experimentally in N-FI-S junctions - demonstrating the ideal half-wave rectification at low frequencies, and  directly probing the second harmonic using lock-in measurements.      

\section{Electronic heat current \label{Heat}}
In this section, we  examine how the e-h symmetry breaking manifests in heat transport in N-FI-S junctions, both in the DC and AC bias. The structure of this section is similar to Sec.~\ref{RectCharge}: first we consider the DC-biased case in Sec.~\ref{SecQDC}, followed by the AC-biased case in Sec.~\ref{SecQAC}. Technical details about the derivation of expressions from this section can be found in Appendix \ref{app1}.
\subsection{DC case \label{SecQDC}}
\begin{figure}[h!]
	\includegraphics[width=0.48\textwidth]{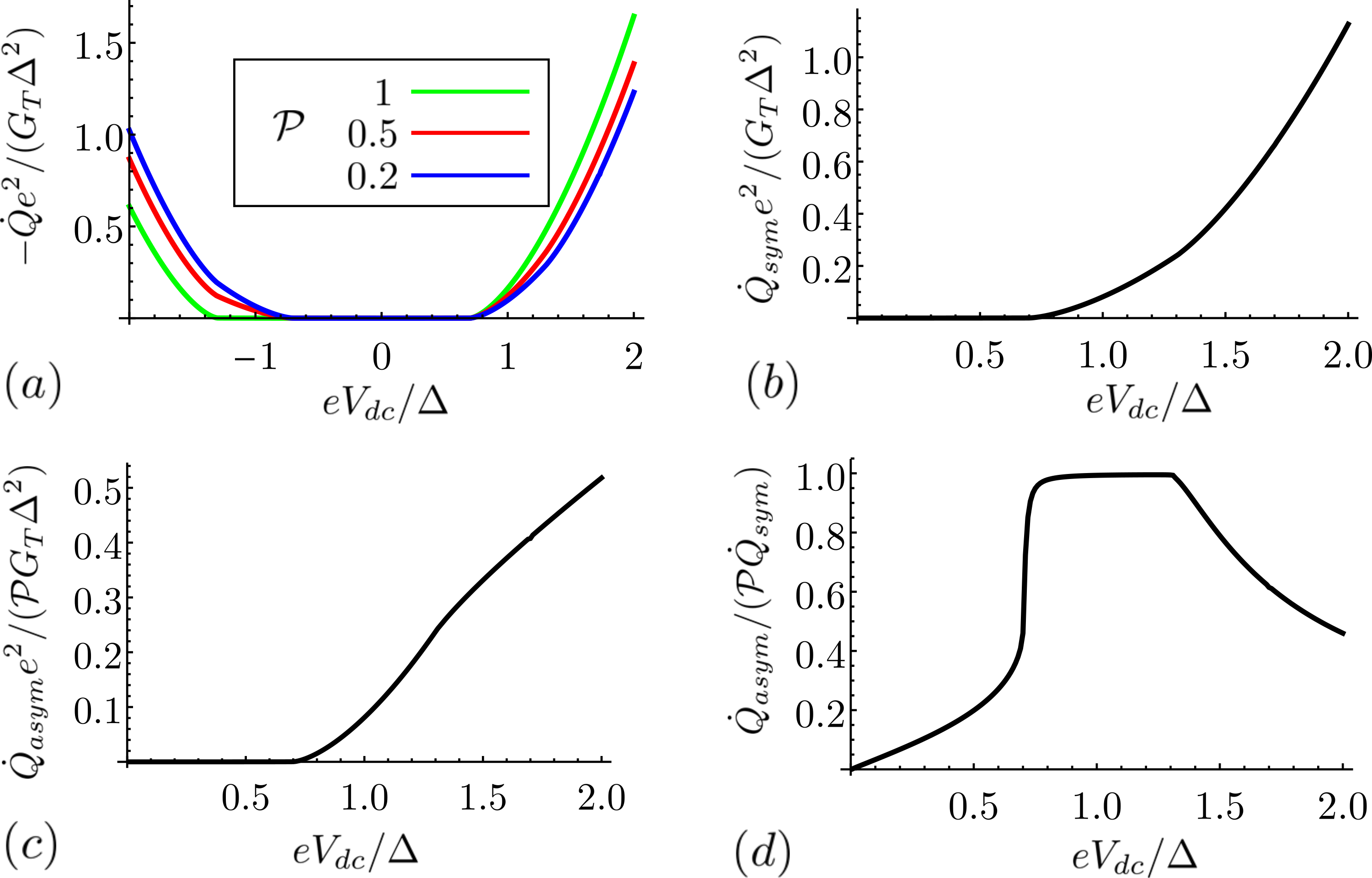}
	\caption{\label{figQDC} Electronic heat current in  N-FI-S junctions with applied DC bias $V_{dc}$ at $T_N=T_S=0$. $(a)$ $\dot{Q}(V)$ curve for different values of the polarization $\mathcal{P}$. $(b)$ Voltage-symmetric component of the heat current. $(c)$ Voltage-asymmetric component of the heat current. $(d)$ Ratio of the symmetric and asymmetric components $\dot{Q}_{asym}/\dot{Q}_{sym}$. The following parameters were used in all plots: $h=0.3 \Delta$, $\eta=10^{-3} \Delta$.  }
\end{figure}

 The electronic heat current flowing to the normal part of the N-FI-S junction is given as
\begin{equation}
	\dot{Q}=\frac{G_T}{e^2}\int dE \tilde{E} [N_++\mathcal{P}N_-]
	[f_N(\tilde{E})-f_S(E)].
	\label{Eq:Qdc}
	\end{equation}
 In Fig.~\ref{figQDC}, we plot  the heat current at low teperatures [panel $(a)$], its symmetric component  $\dot{Q}_{sym}=\frac{1}{2}|\dot{Q}(V_{dc})+\dot{Q}(-V_{dc})|$ [panel $(b)$], and the antisymmetric component $\dot{Q}_{asym}=\frac{1}{2}|\dot{Q}(V_{dc})-\dot{Q}(-V_{dc})|$ [panel $(c)$]. In the absence of polarization, the heat current is symmetric with respect to voltage $\dot{Q}(V)=\dot{Q}(-V)$, but it acquires an antisymmetric component at finite $h$ and $\mathcal{P}$. The ratio $\dot{Q}_{asym}/\dot{Q}_{sym}$ quantifies the asymmetry of the heat current, and it is plotted in  Fig.~\ref{figQDC}$(d)$. Similarly to the results shown in Fig.~\ref{figDC}$(d)$, the maximal asymmetry is achieved in the voltage window $eV_{dc}\in [\Delta-h,\Delta+h]$, and it is determined by the polarization $\mathcal{P}$. 
 
  The existence of a finite $\dot{Q}_{asym}$ leads to an increased cooling of N in N-FI-S junctions at low voltages $V \sim \Delta-h$ \cite{ kolenda2016nonlinear, rouco2018electron}, which is comparable to the cooling found in N-I-S at higher voltages $V \sim \Delta$ \cite{giazotto_opportunities_2006, bergeret2018colloquium}. This effect has a potential to improve cooling for many applications in systems where on-chip electron refrigeration is required. Recently, N-FI-S junctions have also been proposed as efficient thermal rectifiers\cite{giazotto2020very}, {\it i.e.} systems in which the heat current depends on the sign of the thermal gradient across the junction. 
\subsection{AC case \label{SecQAC}}
 As discussed in Sec.~\ref{RectCharge}, e-h symmetry breaking generates a voltage-symmetric component in the current in the DC case, which translates to zero- and even-order harmonics in the AC case. Analogously, voltage-antisymmetric component generated in the heat current in the DC case translate to odd-order harmonics in the AC case, as we discuss in the following. 

 Time-dependent heat current $\dot{Q}(t)$ flowing to the N layer in the AC-biased N-FI-S junction is given as
    \begin{equation}
\dot{Q}(t)=\dot{Q}_0+\sum_{m=1}^{\infty}\bigg[ \dot{Q}_m^c \cos (m  \omega t)+\dot{Q}_m^s \sin (m \omega t)\bigg].
    \end{equation}
 Here, the time-averaged component is
    \begin{equation}
\dot{Q}_0=\frac{G_T}{e^2}\sum_n J_n^2(\beta) \int \text{d}E\,E_n N_+ [f_N(E_n)-f_S(E)],
\end{equation}
and the harmonics are
\begin{multline}
\dot{Q}_m^s=\frac{G_T}{e^2}\int dE \sum_n  \mathcal{L}_{mn}^+(\beta,E) \\
[f_N(E_n)-f_S(E)]
\begin{cases}
N_+ ,   & m\text{ even},\\
 \mathcal{P} N_-,  & m\text{ odd}.
\end{cases}
\end{multline}
\begin{multline}
\dot{Q}_s^m=\frac{-iG_T}{e^2}\int dE \sum_n  \mathcal{L}_{mn}^-(\beta,E) \\
f_N(E_n)
\begin{cases}
M_+ ,   & m\text{ even},\\
 \mathcal{P} M_-,  & m\text{ odd}.
\end{cases}
\label{Eq:Qs}
\end{multline}
Here we have  introduced $\mathcal{L}_{mn}^{\pm}(\beta,E)=J_n(\beta)[E_{n+\frac{m}{2}}J_{n+m}(\beta)\pm E_{n-\frac{m}{2}}J_{n-m}(\beta)]$, with $E_{n\pm \frac{m}{2}}=E+(n\pm \frac{m}{2})\omega$. As mentioned above,  odd-order harmonics $\dot{Q}_{2k+1}^c$ and $\dot{Q}_{2k+1}^s$ exist if both $h$ and $\mathcal{P}$ are finite.  

\begin{figure*}[t!]
	\includegraphics[width=0.9\textwidth]{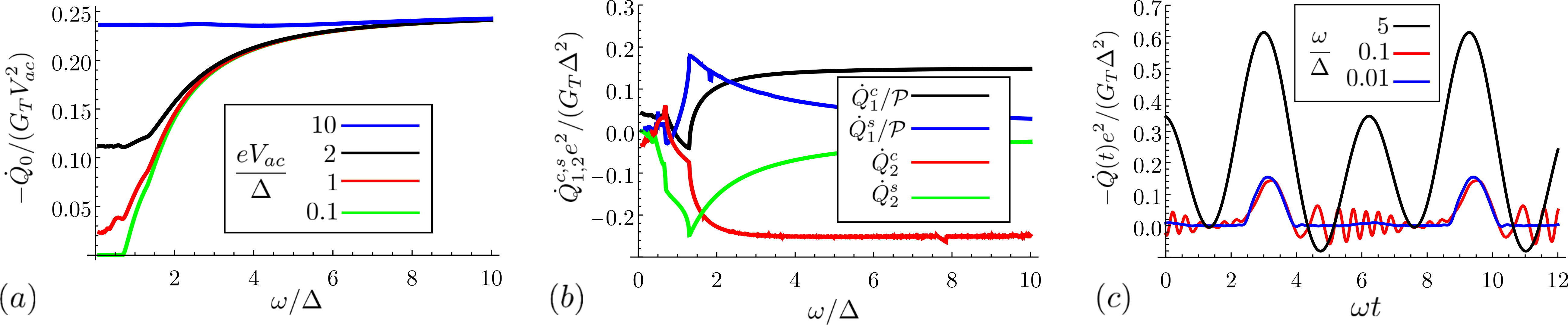}
	\caption{\label{figQAC}Electronic heat current in AC-biased N-FI-S junctions at $T_N=T_S=0$.  In all plots we use the following parameters: $h=0.3 \Delta$, $\eta=10^{-3}\Delta$.  $(a)$ Time-averaged heat current as a function of frequency, at different amplitudes of the bias voltage. $(b)$ First and second harmonic of the heat current as a function of frequency, at $eV_{ac}=\Delta$. $(c)$ Time-dependent heat current, for various values of the frequency, at $\mathcal{P}=0.9$ and $eV_{ac}=\Delta$. Note that we choose a different normalization for heat current in panel $(a)$, and panels $(b)$ and $(c)$ in order to have a better visibility of the different curves.}
\end{figure*}

In Fig.~\ref{figQAC}$(a)$ and $(b)$ we plot $\dot{Q}_0$ and $\dot{Q}_{1,2}^{c}$, respectively, as a function of frequency. In the high frequency limit $\omega \gg \Delta,h,eV_{ac},T$, the $\dot{Q}_0$ and $\dot{Q}_2^c$ saturate to the value
\begin{equation}
\dot{Q}_0=\dot{Q}_2^c\approx -\frac{1}{4}V_{ac}^2G_T,
\label{EqQ0}
\end{equation}
which is the well-known result for Joule heating in AC bias. Note that the prefactor in Eq.~\eqref{EqQ0} is $\frac{1}{4}$ instead of the conventional $\frac{1}{2}$ because we are considering only the heat current flowing to the $N$ electrode.  More interestingly, the first-order harmonic also saturates at high frequencies to a value
\begin{equation}
\dot{Q}_1^c=\frac{\mathcal{P}V_{ac}h}{2e}G_T.
\label{EqQ1}
\end{equation}
Comparing the expressions \eqref{EqQ0} and \eqref{EqQ1}, we see that the first harmonic will actually dominate the time-dependent heat current at high frequency if $2\mathcal{P}h>2 e V_{ac}$. The prominent role of the first harmonic can also be seen in the plot of $\dot{Q}(t)$ in Fig.~\ref{figQAC}$(c)$. 

The time-dependence of heat current, and the peculiar manifestation of the first harmonic, might be difficult to directly probe experimentally. Namely, heat current is usually measured from heat balance using a separately calibrated thermometer, which would need to operate in the adiabatic regime even in the high-frequency limit of the junction. Such a measurement might be feasible with another N-I-S contact with a high-gap superconductor $\Delta>\omega$ . Nevertheless,  the time-dependent heat currents could manifest indirectly in excess noise at low temperatures \cite{averin2010violation}, and also in structures where the heat and charge transport are coupled by the thermoelectric effect, such as detectors.

\section{Analysis of the Noise \label{Noise}}
In this section we turn to noise properties of N-FI-S junctions, in both DC and AC driven regimes. This is of particular interest for application of these junctions as detectors, as the noise sets their performance limits.  As discussed in Appendix \ref{app1}, the heat and charge current presented in the previous sections are calculated  using the Keldysh formalism of non-equlibrium Green's functions. This formalism can be neatly extended to describe the full counting statistics of the junction, including the noise, by using the so-called counting fields \cite{belzig2001full, kindermann2004statistics, belzig2005full}. The details of this approach applied to N-FI-S junctions are described in Appendix \ref{app2}.

Zero-frequency noise correlators for N-FI-S junctions in AC+DC bias are given as
	\begin{align}
	 \langle\delta I^2\rangle&=2 G_T\sum_n \int dE \mathcal{B}_n(E), \nonumber \\
	 \langle\delta I\delta\dot{Q}\rangle&=-\frac{2G_T}{e}\sum_n \int dE \tilde{E}_n \mathcal{B}_n(E),\nonumber \\
	\langle\delta\dot{Q}^2\rangle&=\frac{2G_T}{e^2}\sum_n \int dE \tilde{E}_n^2 \mathcal{B}_n(E).   
	\label{Eq:Noise}
	\end{align}
In addition to usual charge current correlator $\langle\delta I^2\rangle$ and the heat current correlator $\langle\delta\dot{Q}^2\rangle$, the mixed charge-heat current corelator $\langle\delta I\delta\dot{Q}\rangle$ is also finite, which is possible only if e-h symmetry is broken. We introduced $\tilde{E}_n=E+n\omega-eV_{dc}$, and $\mathcal{B}_n(E)= J_n^2(\beta)[N_++\mathcal{P}N_-] \bigg[f_N(\tilde{E}_n)[1-f_S(E)]+f_S(E)[1-f_N(\tilde{E}_n)]\bigg]$. Equation \eqref{Eq:Noise} is a general expression valid at any temperature and bias voltages $V_{ac}$ and $V_{dc}$. In the following we discuss several regimes of interest where significant simplifications are possible.

In the pure thermal regime $(T_N=T_S\gg eV_{ac},eV_{dc})$, we find the well known expression for the thermal noise, namely \cite{golubev2001nonequilibrium, giazotto_opportunities_2006}
	\begin{equation}
	  \langle\delta I^2\rangle=4TG, \quad \langle \delta I\delta\dot{Q}\rangle=-4T\alpha, \quad \langle\delta\dot{Q}^2\rangle=4T^2 G_{th}.
	  \label{Eq:Thermal}
	\end{equation}
Here $G$ is linearized conductance of the junction, $\alpha$ is the thermoelectric coefficient, and $G_{th}$ is the linearized heat conductance. They read
\begin{align}
G&=G_T \int dE \frac{N_+}{4T \cosh^2 (\frac{E}{2T})}, \nonumber \\
G_{th}&=G_T \int dE \frac{E^2N_+}{4T^2 \cosh^2 (\frac{E}{2T})}, \nonumber \\
\alpha&=G_T \mathcal{P}\int dE \frac{E N_-}{4T \cosh^2 (\frac{E}{2T})}.
\label{Eq:G}
\end{align}
These quantities can be approximated analytically in the limit $\Delta-h\gg T$, as detailed in Ref.~\onlinecite{ozaeta_predicted_2014}.

In the DC-biased regime ($V_{dc}\neq 0, V_{ac}=0$), we recover the known relations for the shot noise ($T=0$) \cite{golubev2001nonequilibrium, giazotto_opportunities_2006}
	\begin{equation}
	  \langle\delta I^2\rangle=2e|I|, \quad |\langle \delta I \delta\dot{Q}\rangle|=2 e |\dot{Q}|.
	\end{equation}

\begin{figure*}[t!]
	\includegraphics[width=0.9\textwidth]{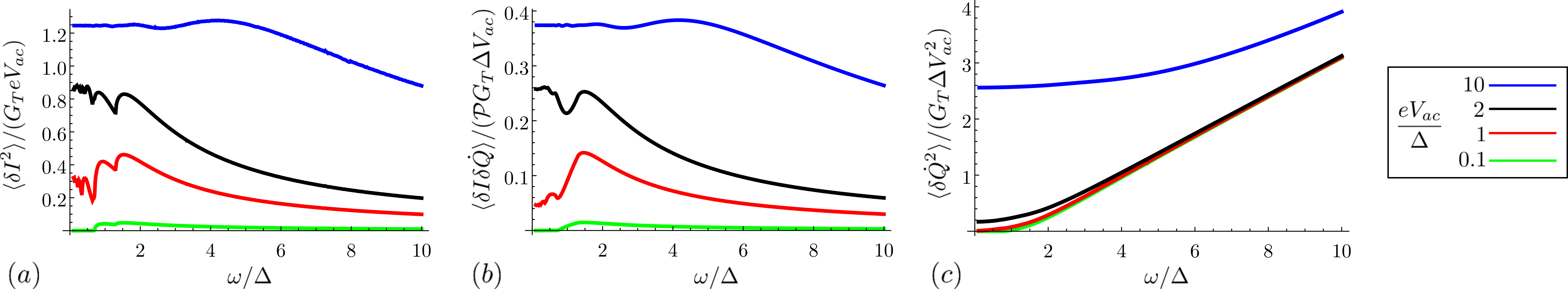}
	\caption{\label{figNoise} AC-driven shot noise correlators as a function of frequency, at different amplitudes of the AC bias voltage. $(a)$ Charge current noise correlator $\langle \delta I^2\rangle$. $(b)$ Mixed heat-charge noise current correlator $\langle \delta I \delta \dot{Q} \rangle$. (c) Heat current noise correlator $\langle \delta Q^2\rangle$. Parameters used in the plots are $h=0.3 \Delta$, $T_N=T_S=0$, $\eta=10^{-3}\Delta$. }
\end{figure*}

Finally, we consider the AC-driven shot noise ($V_{ac}\neq 0, V_{dc}=0, T=0$). The correlators specified in Eq.~\eqref{Eq:Noise} are strongly dependent of frequency, as illustrated in Fig.~\ref{figNoise}. In  the high frequency regime $(\omega\gg \Delta,h,eV_{ac})$, these expressions can be approximated as
	\begin{align}
	   &\langle\delta I^2\rangle=G_T\frac{e^2V_{ac}^2}{2\omega}, \quad
	   \langle\delta I\delta\dot{Q}\rangle=\frac{\mathcal{P} G_T}{e}\frac{e^2V_{ac}^2}{\omega}h,\nonumber \\
	   &\langle \delta \dot{Q}^2\rangle=\frac{2}{3}G_T V_{ac}^2\omega.
	\end{align}
 
 To probe the AC-driven shot noise in experiment, a particularly useful technique is to study the differential noise and its "steps" as functions of an applied DC voltage, as was done in Refs.~\onlinecite{schoelkopf1998observation} and \onlinecite{kozhevnikov2000observation} for a normal conductor and a N-S junction. The differential noise current correlator at zero temperature is given as
\begin{multline}
\frac{\partial \langle\delta I^2\rangle}{\partial V_{dc}}=-G_T\sum_{n} J_n^2(\beta)
\\ [N_+(n\omega-eV_{dc})-\mathcal{P}N_-(n\omega-eV_{dc})]\text{sgn}(n\omega-eV_{dc}).
\end{multline}
\begin{figure}[h!]
	\includegraphics[width=0.275\textwidth]{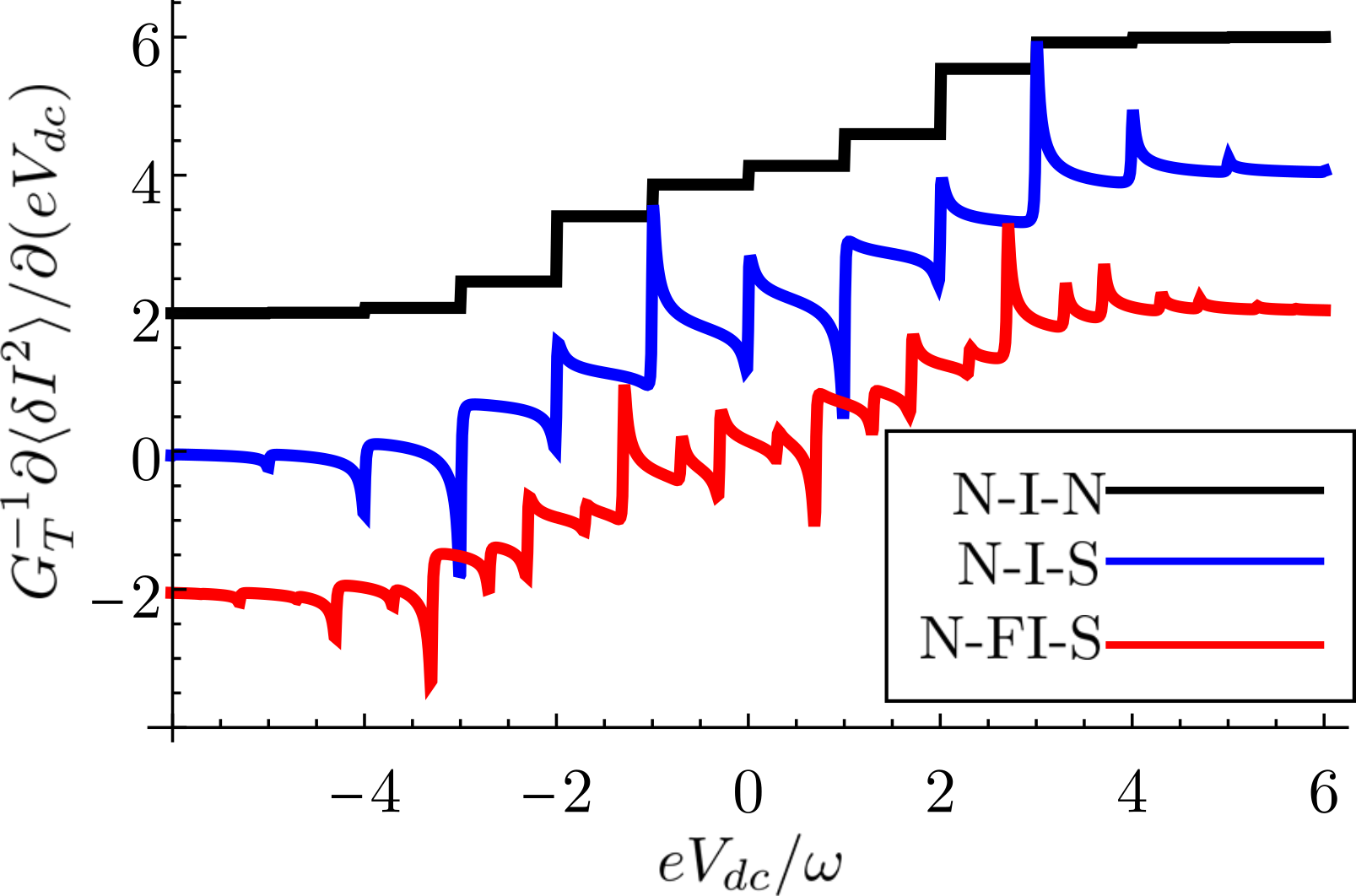}
	\caption{\label{figDifNoise} Comparison of the differential AC+DC-driven shot noise in different tunnel junctions. Black curve describes the normal tunnel junction, blue curve the N-I-S junction, and red curve the N-FI-S junction. Curves are vertically shifted for clarity. In all plots we set $eV_{ac}=3\omega$; for blue curve $\Delta=\omega$; for red curve $\Delta=\omega$, $h=0.3 \Delta$, $\mathcal{P}=0.5$.}
\end{figure}
In a fully normal structure (N-I-N junction), the differential noise has "steps" at $eV_{dc}=n \omega$, where the height of each step is $2G_T J_n^2(\beta)$. Such steps were indeed observed in Ref.~\onlinecite{schoelkopf1998observation}. This is illustrated as a black curve in Fig.~\ref{figDifNoise}. In N-I-S structure, every step acquires a sharp feature due to the non-trivial BCS density of states, as seen in the blue curve in Fig.~\ref{figDifNoise}. In N-FI-S junctions, every step additionally becomes spin-split. Moreover, due to the finite polarization, the curve is no longer fully antisymmetric with respect to voltage. Therefore, it is possible to see strong signature of both spin-splitting and spin-polarization in the differential noise of N-FI-S.

\section{Comparison of the rectification and thermoelectric effect \label{RectvsTherm}}
One of the most striking consequences of e-h symmetry breaking in N-FI-S junctions is a particularly strong thermoelectric effect\cite{ozaeta_predicted_2014}, which can be used as a basis for a radiation detector\cite{heikkila_thermoelectric_2018}. In such detectors, a radiation source heats the junction creating the temperature gradient, which leads to a finite current due to thermoelectric effect. However, such radiation source can also create an alternating voltage along the junction, which will create an additional current due to the rectification effect, as established in Sec.~\ref{SecIAC}. In this section we address several important questions that now arise. First,  how do the rectification and theroelectric current compare? Second,  do the two currents compete or add up? And finally, and how does the presence of rectification affect the radiation detector?

We start from the expression for the rectification and thermoelectric current
\begin{equation}
I_{Rc}=\frac{\mathcal{P} G_T}{2e}\frac{e^2V_{ac}^2}{\omega^2}h, \qquad 
I_{Th}=-\frac{\Delta T}{T} \alpha,
\label{Eq:Detector1}
\end{equation}
 which are valid at high frequencies $\omega\gg \Delta,h,eV_{ac}$, and assuming linear response regime in temperature $\Delta-h\gg T\gg T_S-T_N=\Delta T$. The sign of $I_{Rc}$ is determined by the relative sign of $h$ and $\mathcal{P}$, namely $\text{sign}(I_{Rc})=\text{sign} (h\mathcal{P})$. On the other hand, we know that $\text{sign} (\alpha)=\text{sign} (h \mathcal{P} )$, and therefore $\text{sign} (I_{Th})=\text{sign} (h \mathcal{P} \Delta T)$. This means that $I_{Rc}$ and $I_{Th}$ have the same sign if $\Delta T<0$, therefore, if the normal side of the junction is heated more. On the other hand, if the superconducting side of the junction is heated more, the two currents compete.

\begin{figure*}[t!]
	\includegraphics[width=0.9\textwidth]{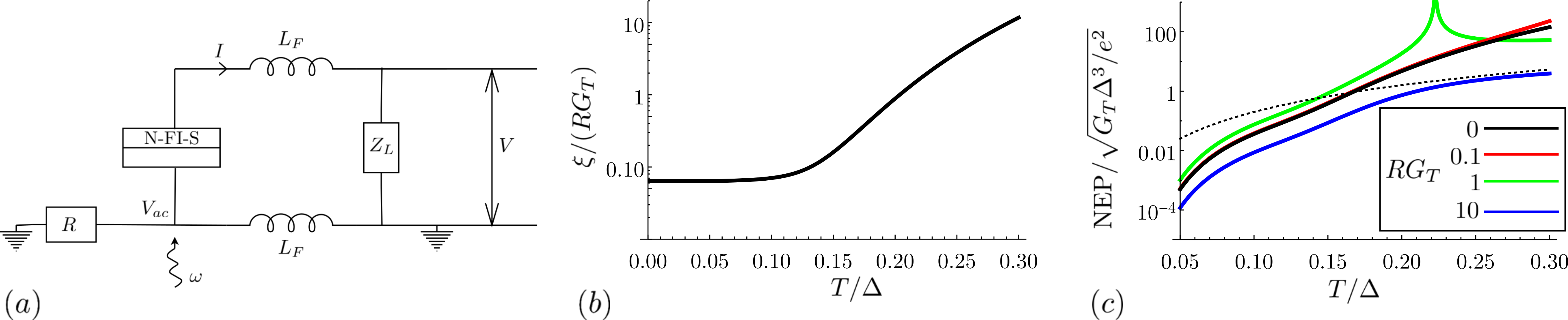} 
	\caption{$(a)$ Schematic representation of the radiation detector based on rectification and thermoelectric effect. Superconducting absorber, whose resistance is $R$, is irradiated with radiation of frequency $\omega$. This creates an AC voltage and a temperature gradient across the N-FI-S junction, which both contribute to a finite current $I$ due to rectification and thermoelectric effect, respectively. Two inductors, $L_F$ serve to prevent any AC currents from reaching the load element, with an impedance $Z_L$. The readout voltage of the detector is denoted as $V$.  $(b)$ Plot of the parameter $\xi$, which describes the relative strength of the rectification effect in comparison with the thermoelectric effect,  as a function of temperature. $(c)$ NEP of the detector as a function of temperature, for different absorber resistances $R$. Black dotted line corresponds to the NEP of a hot electron bolometer, given as NEP=$\sqrt{20\Sigma \Omega T^6}$. In panels $(b)$ and $(c)$, we used the following parameters: $\Sigma\Omega \Delta^3 e^2=0.2 10^4 G_T$, $h=0.2 \Delta$, $\omega=3\Delta$, $\mathcal{P}=0.5$. \label{Fig:Detector}}
\end{figure*}

Next, we take that the N-FI-S junction is a part of the radiation detector circuit shown in Fig.~\ref{Fig:Detector}$(a)$. Here, the superconducting absorber is irradiated with high-frequency radiation ($\omega\gg \Delta$), leading both to the temperature gradient $\Delta T$ and an AC bias voltage $V_{ac}$ across the junction. As in this setup the superonductor heats more than the normal side of the junction, thermoelectric and rectification effect will compete.   
The detector current $I$ can be written as
\begin{equation}
I=\alpha \frac{\Delta T}{T}-GV-\frac{\mathcal{P}G_T}{2e}\frac{e^2V_{ac}^2}{\omega^2}h,
\label{Eq:Detector2}
\end{equation}
where $V$ is the readout voltage, which can be expressed also as $V=I Z_L$, and $Z_L$ is the load impedance.

In order to determine $\Delta T$, we write the heat balance equation:
\begin{equation}
P_\gamma=G_{th}^{tot}\Delta T-\alpha V+\frac{V_{ac}^2}{4}G_T.
\label{Eq:Detector3}
\end{equation}
Here $P_\gamma=V_{ac}^2/(2R)$ is the total incoming power, where $R$ is the absorber resistance.  $G_{th}^{tot}=G_{th}+G_{e-ph}$ is the total heat conductance, where $G_{th}$ and $G_{e-ph}$ are heat conductances from quasiparticles to the normal side of the junction, and from quasiparticles to the phonons, respectively. The term $\alpha V$ describes the Peltier heat current driven by the voltage $V$. The last term in Eq.~\eqref{Eq:Detector2} stems from the Joule heating of the junction.

The electron-phonon heat conductance for the superconducting absorber is given as\cite{virtanen2016stimulated, heikkila_thermoelectric_2018}
\begin{equation}
G_{q-ph}=
\Sigma \Omega T^4 \Phi(\Delta,h,T),
\end{equation}
where $\Omega$ is the volume of the absorber and $\Sigma$ is material-dependent electron-phonon coupling constant. The function $\Phi$ for $T\ll\Delta-|h|$ is given as $\Phi= \frac1{96 \zeta(5)}[f_1(\tilde{\Delta}) \cosh \tilde{h}e^{-\tilde{\Delta}} +\pi \tilde{\Delta}^5 f_2(\tilde{\Delta})e^{-2\tilde{\Delta}}]$,
where $\tilde{\Delta}=\frac{\Delta}{T}$, $\tilde{h}=\frac{h}{T}$. Moreover, we introduced $f_1(x)=\sum_{n=0}^3 C_n/x^n$, $f_2(x)=\sum_{n=0}^{2}B_n/x^n$, with $C_0\approx 440$, $C_1\approx 500$, $C_2\approx 1400$, $C_3\approx 4700$, $B_0=64$, $B_1\approx 144$ and $B_2\approx 258$.

Solving Eqs.~\eqref{Eq:Detector1}-\eqref{Eq:Detector3} we find the voltage $V$, and calculate the so-called voltage responsivity $\lambda_V$ as
\begin{equation}
\lambda_V=\frac{V}{P_\gamma}=\bigg(1-\frac{RG_T}{2}-\xi\bigg) \lambda_V^0.
\end{equation}
Here, $\lambda_V^0=\alpha/[G_{th}^{tot}Y_{tot}T-\alpha^2]$ is the responsivity in the absence of the rectification effect \cite{heikkila_thermoelectric_2018}, with $Y_{tot}=G+Z_L^{-1}$ being the total electrical admittance. The parameter 
\begin{equation}
\xi=e\mathcal{P} R G_T \frac{hT}{\omega^2} \frac{ G_{th}^{tot}}{\alpha}
\end{equation}
determines the relative strength of rectification effect in comparison to thermoelectric effect - at $\xi=0$ the rectification effect vanishes, whereas $\xi\gg 1$ corresponds to a very strong rectification effect. We plot this parameter as the function of temperature in Fig.~\ref{Fig:Detector}$(b)$. In the case when electron-phonon coupling gives the dominant contribution to the heat conductance,  $G_{q-ph}\gg G_{th}$, we can approximate 
\begin{equation}
\xi\approx \frac{e \Sigma h T^5 \Phi}{\omega^2 \tilde{\alpha}}\rho L^2
\end{equation}
Here, we introduced the specific resistance of the absorber $\rho$, and the length of the absorber $L$, and $\tilde{\alpha}=\alpha/(\mathcal{P}G_T)$. Therefore, the relative strength of rectification and thermoelectric effect strongly depends on the length of the junction, with the former being more pronounced in longer junctions.

Next, we turn to the noise properties of the detector.  We write the Kirchof's law for the noise terms
\begin{align}
G_{th}^{tot}\delta T&=\delta \dot{Q}_{q-ph}+\delta\dot{Q}+\alpha \delta V, \nonumber \\
Y_{tot}\delta V&=\delta I+\alpha \delta T/T.
\label{Eq:KirNoise}
\end{align}
Here, $\delta T$ describes the temperature fluctuations, and we introduced  the noise due to electron-phonon coupling $\langle \delta \dot{Q}^2_{q-ph}\rangle=4T^2 G_{q-ph}$. In the following, we assume that the temperature is high enough, so that thermal noise dominates and shot noise can be neglected. In this case, we can use the noise correlators specified in Eq.~\eqref{Eq:Thermal}. Then, from Eq.~\eqref{Eq:KirNoise} we find
\begin{equation}
\langle \delta V^2\rangle=|\lambda_V^0|^2 \frac{T^2}{\alpha^2} G_{th}^{tot} (G T G_{th}^{tot}-\alpha^2).
\end{equation}

Finally, we calculate the noise-equivalent power (NEP) of the detector $P_{ne}$, which quantifies its noise-to-signal ratio
\begin{equation}
P_{ne}=\frac{\sqrt{\langle \delta V ^2\rangle}}{\lambda_V}=\frac{P_{ne}^0}{1-\frac{RG_T}{2}-\xi}.
\label{Eq:NEP}
\end{equation}
Here
\begin{equation}
P_{ne}^0=\frac{2T}{\alpha}\sqrt{G_{th}^{tot}(G T G_{th}^{tot}-\alpha^2)},
\end{equation}
 is the NEP of the detector in the absence of rectification effect, as established in Ref.~\onlinecite{heikkila_thermoelectric_2018}.
 
 There are three conductances that enter Eq.~\eqref{Eq:NEP}: $G_T$, $1/R$ and $\Sigma \Omega \Delta^3e^2$. In the following, we set $G_T=5 \times 10^{-4} \Sigma \Omega \Delta^3 e^2$, which was estimated to be optimal for thermoelectric effect in Ref.~\onlinecite{heikkila_thermoelectric_2018}. In Fig.~\ref{Fig:Detector}$(c)$, we plot the NEP of the detector for different values of $RG_T$. For comparison, in the same plot we include the NEP of a bolometer of the same volume, whose heat conductance is limited by electron-phonon coupling, so that NEP=$\sqrt{20\Sigma \Omega T^6}$ \cite{karasik2011demonstration}. At $RG_T=0$, we recover the NEP of the purely thermoelectric detector. Increasing $RG_T$ to 0.1 has a very slight effect on NEP. The worst performance is found at $R G_T=1$ - in this case thermoelectric and rectification effect can completely cancel out, leading to a vanishing signal and a divergence in the NEP. Finally, at $RG_T=10$ the NEP significantly improves.
 
 These results show that the detector has a good perfomance if either rectification or thermoelectric effect dominates. However, if they are of comparable strength, they compete and performance is at worst. Therefore, special care must be taken when building the detector to optimize the ratio of these two effects, which depends on material-specific parameters such as $\rho$ and $\Sigma$, the temperature and the frequency of radiation,  as well as length of the junction.

\section{Rectification of environmental noise\label{DCB}}
 In Sec.~\ref{RectCharge}, we demonstrated that AC signals are rectified in N-FI-S junctions, but we considered only an ideal situation in the absence of an electromagnetic environment. Such environments can act as sources of noise, which can also be rectified by the junction and lead to a finite current even in the absence of applied voltages and temperature gradients. A directly applicable tool to study this noise is the $P(E)$ theory of Coulomb blockade \cite{ingold1992charge}. Namely, in the presence of the noise described by $P_n(E)$, the tunneling current expression becomes
\begin{widetext}
\begin{equation}
I=\frac{G_T}{e}\int d E d E' [N_+ (E)+\mathcal{P}N_-(E)] \\
\bigg[
P_n(\tilde{E}'-E)f_N(\tilde{E}')[1-f_S(E)]-P_n(E-\tilde{E}')f_S(E)[1-f_N(\tilde{E}')]
\bigg],
\end{equation}
\end{widetext}
where we introduced $\tilde{E}'=E'-eV_{dc}$. For an ideal environment we have $P_n(E-E')=\delta(E-E')$, in which case we recover Eq.~\eqref{eqIDC}.

\begin{figure}[h!]
	\includegraphics[width=0.3\textwidth]{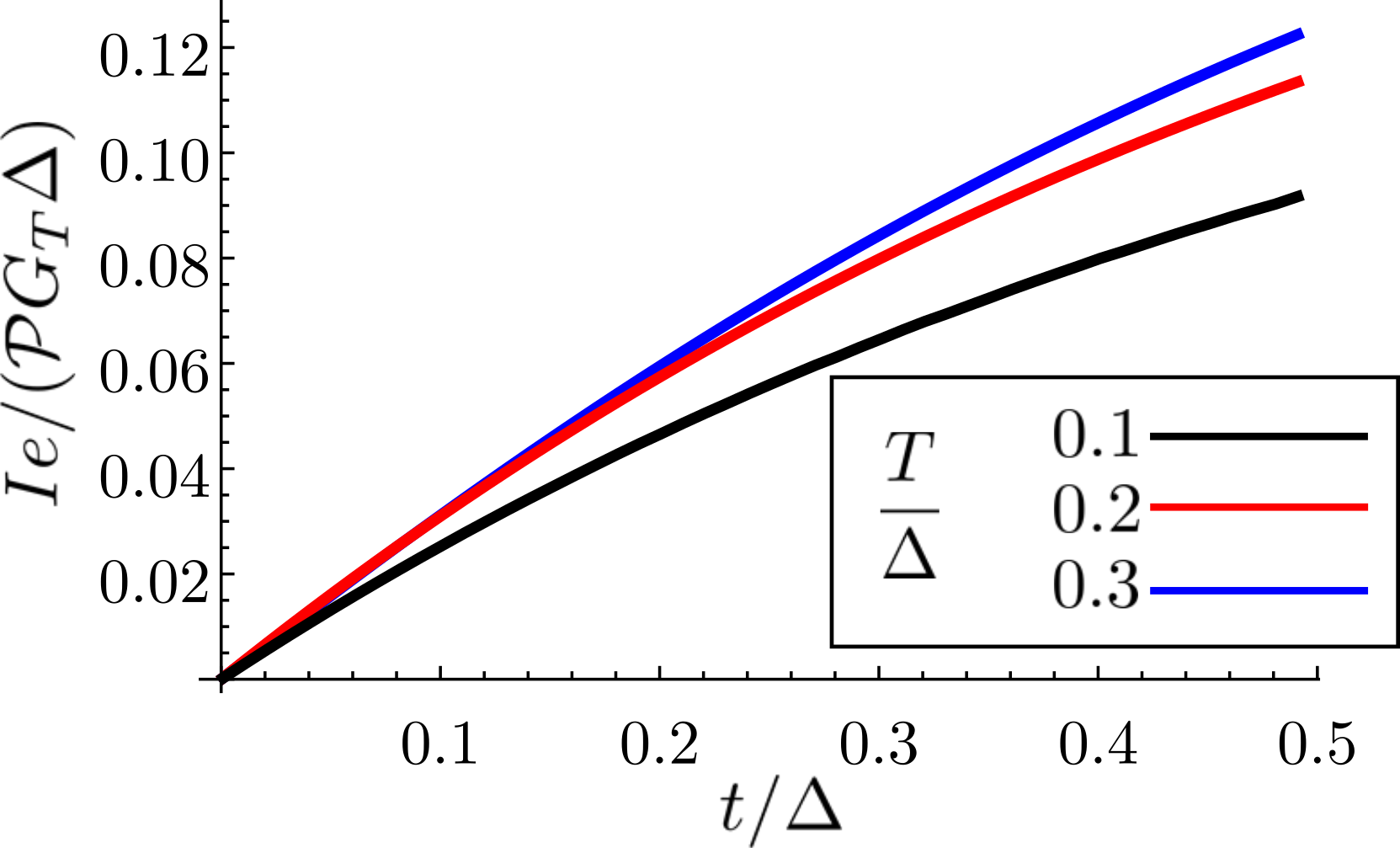}
	\caption{Current resulting from the rectification of environmental noise, as a function of the environment parameter $t$, for different temperatures $T=T_N=T_S$. We used $h=0.5 \Delta$ and $\eta=10^{-3}\Delta$. \label{Fig:Ivst}}
\end{figure}

One possibility is to describe the $P_n(E)$ due to a high-temperature environment, as in Ref.~\onlinecite{pekola2010environment}. There, for $E\ll T_{env}\hbar /(R_{env}C_{env})$, the authors derive a very simple Lorentzian function
\begin{equation}
  P_n(E)=\frac{1}{\pi} \frac{t}{t^2+\epsilon^2},
\end{equation}
where $t=R_{env} T_{\rm env}/R_Q$ describes the noise source with
resistance $R_{env}$ at temperature $T_{\rm env}$ and $R_Q=\hbar/e^2$ is the
resistance quantum (note $\hbar$ instead of $h$).
The presence of such a noise source leads to a finite current even at
a vanishing voltage or temperature difference across the
junction:
\begin{equation}
I=\frac{\mathcal{P}G_T}{e}\int dE dE' N_-(E) P_n(E-E')[f_N(E')-f_S(E)].
\end{equation}
This finite current is plotted in Fig.~\ref{Fig:Ivst}
as a function of the parameter $t$. Notably, the current
is a decreasing function of temperature.

\section{Conclusion \label{Conclusion}}
Electron-hole symmetry breaking occurs in N-FI-S junctions due to the combination of spin-splitting and spin-filtering. As a consequence, N-FI-S junctions act as rectifiers of charge current, and could therefore present a promising platform to be utilized as a diode for emerging low-temperature electronics, or in radiation detectors.  The crucial parameter determining the efficiency of rectification is the spin-polarization of the tunneling barrier $\mathcal{P}$. Importantly, rectification effect might be in competition with a  thermoelectric effect, and the two effects need to be considered simultaneously in realistic structures such as detectors. The interplay of the two effects  can be optimized by varying the size of the junction and the choice of materials. A first confirmation of the rectification effect discussed in the present work has been recently reported in Ref. \cite{strambini2021rectification} for a Al/EuS/Cu junction.

Our work also provides a full characterization of heat transport and noise in N-FI-S junctions, for both AC and DC bias, as well as analysis of the influence of an electromagnetic environment. All of these findings are important to understand and achieve optimal performance  of N-FI-S junctions in numerous applications, such as radiation detection\cite{heikkila_thermoelectric_2018}, electron refrigeration\cite{rouco2018electron} and thermometry\cite{giazotto_ferromagnetic-insulator-based_2015}.   

\acknowledgments

We thank Elia Strambini and Francesco Giazotto for discussions. This project has received funding from the European Union’s Horizon 2020 research and innovation program under grant agreement No. 800923 (SUPERTED). It was also supported by the Academy of Finland (project number 317118), and the Spanish Ministerio de Ciencia e Innovación  through project PID2020-114252GB-I00 (SPIRIT).

\appendix 
\section{Derivation of  $I(t)$ and $\dot{Q}(t)$ from non-equilibrium Keldysh Green's functions\label{app1}}
Within the non-equilibrium theory of Keldysh Green's functions (GFs), the charge current through a N-FI-S junction biased by the time-dependent voltage $V(t)=-V_{dc}+V_{ac}\cos \omega t$ can be expressed as \cite{bergeret2012electronic}
	\begin{equation}
	I(t)=\frac{G_T\pi}{16 e}\text{Tr}\tau_z\eta_x [\Gamma \check{g}_N\Gamma,\check{g}_S]_\circ(t,t')|_{t=t'}.
	\label{eqA1}
	\end{equation}
The electronic heat current is \cite{kopnin2008influence}
	\begin{multline}
	 \dot{Q}(t)=\frac{G_T\pi}{16 e^2}\text{Tr}\eta_x (i\partial_t-i\partial_{t'}+2eV(t)\tau_z) \\ [\Gamma \check{g}_N\Gamma,\check{g}_S]_\circ(t,t')|_{t=t'}.
	 \label{eqA2}
	\end{multline}
Here, $\check{g}_N(t,t')$ and $\check{g}_S(t,t')$ are the two-time GFs for the normal and superconducting side of the junction, respectively. $\check{g}_{N,S}$ are matrices in spin, Nambu and Keldysh space, spanned by the Pauli matrices $\sigma_i$, $\tau_i$, and $\eta_i$ $(i=x,y,z)$, respectively. The symbol $\circ$ denotes a convolution over the intermediate time variable. The spin-polarization of the junction is described by the matrix $\Gamma=t+u\tau_z\sigma_z$, with $t^2+u^2=1$ and $2tu=\mathcal{P}$.

The non-equilibrium GFs, $\check{g}_{N,S}(t,t')$, can be expressed in terms of equilibrium ones, $\check{g}_{N,S}(t-t')$, as  
 \begin{align}
	&\check{g}_N(t,t')=e^{i \phi(t) \tau_z} \check{g}_N(t-t')\nonumber e^{-i\phi(t')\tau_z},\\  &\check{g}_S(t,t')=\check{g}_S(t-t').
	\label{eqA3}
	\end{align}
Here, the bias voltage is introduced via the gauge transformation of the normal-side GF, with $\phi(t)=-V_{dc} t+\beta \sin \omega t$, and $\beta=e V_{ac}/\omega$. Next, the equilibrium GFs can be expressed in energy representation as
	\begin{equation}
	\check{g}_{N,S}(t-t')=\int\frac{dE}{2\pi} e^{-i E(t-t')}\check{g}_{N,S}(E).
	\label{eqA4}
	\end{equation}   
From here, using the Jacobi-Anger expansion, we may write the normal-side GF as
\begin{multline}
\check{g}_N(t,t')=\sum_{nn'} J_n(\tau_z \beta) J_{n'}(\tau_z \beta) \\
\int \frac{d E}{2\pi} e^{-i (t-t')(E+V_{dc}\tau_z)} e^{i\omega (tn-t'n')}\check{g}_N(E).
\label{eqA5}
\end{multline}
The GFs have a following structure in the Keldysh space
\begin{equation}
	\check{g}_{N,S}(E)=\begin{pmatrix}
	g^R_{N,S}(E) & g^K_{N,S}(E) \\
	0 & g^A_{N,S}(E)
	\end{pmatrix},
	\label{eqA6}
	\end{equation}
where the superscripts $R,A$ and $K$ stand for retarded, advanced and Keldysh component, respectively. They are given as $g_N^{R,A}(E)=\pm \tau_z$, $g_N^K=2\tau_z \tanh{E/2T_N}$,  $g_{S\uparrow, \downarrow}^{R,A}(E)=\tau_z(E\pm h\pm i\eta)/\xi_{\uparrow,\downarrow}^{R,A}+i\tau_y\Delta/\xi_{\uparrow,\downarrow}^{R,A}$, $g_{S\uparrow,\downarrow}^{K}(E)=[g^{R}_{S\uparrow,\downarrow}(E)-g^{A}_{S\uparrow,\downarrow}(E)]\tanh{E/2T_S}$. Here $\xi^{R,A}_{\uparrow, \downarrow}=\sqrt{(E\pm h\pm i\eta)^2-\Delta^2}$.

First, we calculate the charge current $I(t)$ in the AC+DC bias. Combining Eqs.~\eqref{eqA3}-\eqref{eqA6} with Eq.~\eqref{eqA1}, we find $I(t)$ as a sum over harmonics
 \begin{equation}
I(t)=I_0+\sum_{m=1}^{\infty}\bigg[ I_m^c \cos (m  \omega t)+I_m^s \sin (m \omega t)\bigg],
\label{eqA7}
\end{equation}
with 
\begin{multline}
I_m^c=\frac{G_T}{e}\sum_n \mathcal{M}_{mn}^+(\beta) \\
\int dE\begin{cases}
 \mathcal{P} N_- [f_N^+(E_n)-f_S(E)]+N_+f_N^-(E_n) \quad m\text{ even},\\
 N_+ [f_N^+(E_n)-f_S(E)]+\mathcal{P}N_-f_N^-(E_n)  \quad m\text{ odd},
\end{cases}
\label{eqA8}
\end{multline}
\begin{multline}
I_m^s=\frac{-iG_T}{e}\sum_n \mathcal{M}_{mn}^{-}(\beta) \\
\int dE\begin{cases}
 \mathcal{P} M_- f_N^+(E_n)+M_+f_N^-(E_n) \quad m\text{ even},\\
 M_+ f_N^+(E_n)+\mathcal{P}M_-f_N^-(E_n)  \quad m\text{ odd}.
\end{cases}
\label{eqA9}
\end{multline}
Here we use the notation $E_n=E+n\omega$, we introduced $f_N^{\pm}(E)=\frac{1}{2}[f_N(E-eV_{dc})\pm f_N(E+eV_{dc})]$, and $\mathcal{M}_{nm}^{\pm}$ is introduced below Eq.~\eqref{Eq:Is} of the main text. The time-averaged component of the current is $I_0=\frac{1}{2}I_c^0$. Setting $V_{dc}\neq0, V_{ac}=0$ (DC regime) or $V_{dc}=0, V_{ac}\neq 0$ (AC regime) in Eqs.~\eqref{eqA7}-\eqref{eqA9}, we obtain the results presented in Secs.~\ref{SecIDC} and \ref{SecIAC}, respectively.  

Next, we calculate the electronic heat current $\dot{Q}(t)$ in the AC+DC bias.  Combining Eqs.~\eqref{eqA3}-\eqref{eqA6} with Eq.~\eqref{eqA2}, we obtain
\begin{equation}
\dot{Q}(t)=\dot{Q}_0+\sum_{m=1}^{\infty}\bigg[ \dot{Q}_m^c \cos (m  \omega t)+\dot{Q}_m^s \sin (m \omega t)\bigg],
\label{eqA10}
\end{equation}
where
\begin{multline}
\dot{Q}_m^c=\frac{G_T}{e^2}\sum_n \int dE \mathcal{L}_{mn}^+(\beta,E) \\
\begin{cases}
N_+ [f_N^+(E_n)-f_S(E)]+\mathcal{P}N_-f_N^-(E_n)  \quad m\text{ even}.\\
 \mathcal{P} N_- [f_N^+(E_n)-f_S(E)]+N_+f_N^-(E_n) \quad m\text{ odd},
\end{cases}
\label{eqA11}
\end{multline}
\begin{multline}
\dot{Q}_m^s=\frac{-iG_T}{e^2}\sum_n \int dE \mathcal{L}_{mn}^-(\beta,E)\\
\begin{cases}
M_+ f_N^+(E_n)+\mathcal{P}M_-f_N^-(E_n)  \quad m\text{ even}.\\
 \mathcal{P} M_- f_N^+(E_n)+M_+f_N^-(E_n) \quad m\text{ odd}.
\end{cases}
\label{eqA12}
\end{multline}
The quantity $\mathcal{L}_{mn}^{\pm}$ is introduced below Eq.~\eqref{Eq:Qs}. The time-averaged component of the heat current is $\dot{Q}_0=\frac{1}{2}\dot{Q}_c^0.$  Setting $V_{dc}\neq0, V_{ac}=0$ (DC regime) or $V_{dc}=0, V_{ac}\neq 0$ (AC regime) in Eqs.~\eqref{eqA10}-\eqref{eqA12}, we obtain the results presented in Secs.~\ref{SecQDC} and \ref{SecQAC}, respectively.

\section{Derivation of the noise correlators \label{app2}}
Refs.~\onlinecite{belzig2001full, kindermann2004statistics} established a way to calculate the full counting statistics for charge and heat transport in junctions in the language of Keldysh GFs. The main quantity of interest is the generating function $\mathcal{S}$, which can be used to calculate all quantities accessible in transport measurements, such as currents, noise, and higher-order correlators. The time-averaged generating function is given as
\begin{multline}
\mathcal{S}(\chi,\psi)=\frac{G_T}{16 e^2}\sum_n J_n^2(\beta) \int \frac{dE}{2\pi}\text{Tr} \\  \{\Gamma \check{g}_N^{\chi,\psi}(E+n\omega-eV_{dc}\tau_z) \Gamma,\check{g}_S(E) \}.
\label{eqB1}
\end{multline}
The charge and heat current counting fields, $\chi$ and $\psi$, appear in the gauge transformation of the normal-side GF  
\begin{equation}
\check{g}_N^{\chi,\psi}(E)=e^{\frac{i}{2}e\chi\eta_x\tau_z}e^{\frac{i}{2}E\psi\eta_x}\check{g}_N(E) e^{-\frac{i}{2}e\chi\eta_x\tau_z}e^{-\frac{i}{2}E\psi\eta_x}.
\end{equation}
Note that Eq.~\eqref{eqB1} generalizes previous expressions found in the tunneling limit in Refs.~\onlinecite{belzig2001full, kindermann2004statistics} by allowing for a spin-polarized tunneling barrier, described by spin-space matrix $\Gamma$. 

The first moments of the generating function, given by the first derivatives with respect to the relevant counting field, yield the time-averaged charge and heat currents:
\begin{equation}
I_0=\frac{\partial \mathcal{S}}{\partial \chi} \bigg|_{\chi,\psi=0}, \quad \dot{Q}_0=\frac{\partial \mathcal{S}}{\partial \psi} \bigg|_{\chi,\psi=0}.
\end{equation}
One can readily check that Eqs.~\eqref{eqIDC}, \eqref{Eq:I0}, \eqref{Eq:Qdc}, and \eqref{EqQ0} from the main text can be reproduced using this approach. 

The second moments of the generating function are related to noise correlators:
\begin{align}
&\langle\delta I^2\rangle=-2\frac{\partial^2\mathcal{S}}{\partial\chi^2}\bigg|_{\chi,\psi=0}, \quad \langle\delta I \delta\dot{Q}\rangle=2\frac{\partial^2\mathcal{S}}{\partial\chi\partial\psi}\bigg|_{\chi,\psi=0}, \nonumber \\
&\langle\delta \dot{Q}^2\rangle=-2\frac{\partial^2\mathcal{S}}{\partial\psi^2}\bigg|_{\chi,\psi=0},
\end{align}
which yield Eq.~\eqref{Eq:Noise} of the main text. 
\end{document}